\documentclass[%
reprint,
superscriptaddress,
 amsmath,amssymb,
 aps,
prb,
longbibliography,
 lengthcheck,%
]{revtex4-1}

\usepackage{graphicx}
\usepackage{dcolumn}
\usepackage{bm}
\usepackage{hyperref}
\usepackage{sidecap}
\usepackage[mathlines]{lineno}
\makeatletter
\newcommand*{\rom}[1]{\expandafter\@slowromancap\romannumerial#1@}
\makeatother

\begin{document}

\preprint{APS/123-QED}

\title{Investigation of infrared phonon modes in multiferroic single-crystal FeTe$_{2}$O$_{5}$Br}
\author{K. H. Miller}
\affiliation{Department of Physics, University of Florida, Gainesville, Florida 32611-8440, USA}
\author{X. S. Xu}
\affiliation{Materials Science and Technology
Division, Oak Ridge National Laboratory, Oak Ridge, Tennessee 37831, USA}
\author{H. Berger}
\affiliation{Institute of Physics of Complex Matter, Ecole Polytechnique Federal de Lausanne, CH-1015 Lausanne, Switzerland}
\author{V. Craciun}
\affiliation{Major Analytical Instrumentation Center, Materials Science and Engineering, University of Florida, Gainesville, Florida 32611, USA}
\author{X. Xi}
\affiliation{Department of Physics, University of Florida, Gainesville, Florida 32611-8440, USA}
\author{C. Martin}
\affiliation{Department of Physics, University of Florida, Gainesville, Florida 32611-8440, USA}
\author{G. L. Carr}
\affiliation{Photon Sciences, Brookhaven National Laboratory, Upton, New York 11973, USA}
\author{D. B. Tanner}
\affiliation{Department of Physics, University of Florida, Gainesville, Florida 32611-8440, USA}

\begin{abstract}
Reflection and transmission as a function of temperature (5--300~K) have been measured on single crystals of the multiferroic compound FeTe$_{2}$O$_{5}$Br utilizing light spanning the far infrared to the visible portions of the electromagnetic spectrum.  The complex dielectric function and optical properties were obtained via Kramers-Kronig analysis and by fits to a Drude-Lortentz model. Analysis of the anisotropic excitation spectra via Drude-Lorentz fitting and lattice dynamical calculations have lead to the observation of all 52 IR-active modes predicted in the $ac$ plane and 43 or the 53 modes predicted along the $b$ axis of the monoclinic cell.  Assignments to groups (clusters) of phonons have been made and trends within them are discussed in light of our calculated displacement patterns.

\end{abstract}

\maketitle

\section{Introduction}
Recently, significant attention has been given to a class of multiferroics in which specific magnetic ordering breaks inversion symmetry, thus allowing spontaneous polarization to arise.  This magnetically driven ferroelectric response is the most direct form of magnetoelectric coupling and is associated with large magnetoelectric\cite{PRB.82.060402} and magnetocapicative\cite{PRL.92.257201} responses.  Electric dipoles induced by the ordering of magnetic dipoles remain highly susceptible to applied magnetic fields, thus making this class of multiferroics of interest from a technological point of view.\cite{NatMater.6.13}  Recent studies on materials such as TbMnO$_{3}$,\cite{Nature.426.55} RbFe(MoO$_{4}$)$_{2}$,\cite{PRL.98.267205} and CoCrO$_{4}$\cite{PRL.96.207204} all show the appearance of spontaneous polarization due to non-collinear magnetic ordering via spin-orbit mechanisms, viz., the standard spin-current and inverse Dzyaloshinski-Moriya models.  A technological shortcoming of magnetically driven ferroelectrics has been the small value of induced spontaneous electric polarization observed.  Much larger spontaneous polarizations have been predicted in materials where ferroelectricity arises from collinear magnetic ordering of spins.\cite{PRL.97.227204,PRL.99.227201}  In collinear magnetic ordering, moments may vary in amplitude but do not vary in direction; therefore, one does not expect the aforementioned spin-orbit-related interactions to be important,\cite{PRB.79.020404} enabling the much stronger exchange striction interaction as the mechanism by which ferroelectricity is induced.

In this paper, we report our infrared studies on single crystals of the multiferroic FeTe$_{2}$O$_{5}$Br compound, a magnetically driven ferroelectric with nearly-collinear incommensurate spin order.  A previous study by Pregelj {\it et al.\/}\cite{PRB.82.144438} has reported significant changes in the magnetic and electric orders when external magnetic fields are applied along the different crystallographic directions.  Strikingly, for B$\parallel${\it b\/} they observed that fields greater than 4.5~T destroyed the electric polarization completely.  In our investigation of magnetoelectric coupling we measured transmission in the far-infrared down to 15~cm$^{-1}$ (2~meV) and in external  magnetic fields up to 10~T oriented along all three crystallographic axes.  The significant magnetoelectric coupling previously shown by the application of external magnetic fields did not surface in the far infrared to the extent we examined it.  However, we present a comprehensive report of the excitation spectrum in single crystal FeTe$_{2}$O$_{5}$Br.  The modeling of the infrared active phonons as well as lattice dynamical calculations have lead us to make an interesting comparison to a previous infrared study on a similar  geometrically frustrated spin-cluster oxyhalide compound, FeTe$_{2}$O$_{5}$Cl,\cite{JCMP.21.375401} where an equal number of modes was predicted but a significantly lower number was reported.
   
Becker {\it et al.\/}\cite{JACS.128.15469} solved the crystal structure using single-crystal x-ray diffraction.  The FeTe$_{2}$O$_{5}$Br compound crystallizes in a layered structure where the individual layers comprise the {\it bc\/} plane in the monoclinic system (P2$_{1}$/{\it c\/}).  The layers are weakly connected via van der Waals forces and slip at an 18$^{\circ}$ angle as they stack along the normal to the {\it bc\/} plane, thus defining the monoclinc {\it a\/} axis.  A single layer is composed of [Fe$_{4}$O$_{16}$]$^{20-}$ groups confined on top and bottom by [Te$_{4}$O$_{10}$Br$_{2}$]$^{6-}$ groups.  The three groups contain common oxygens which connect them and also serve to create charge neutrality within the layers.  The [Fe$_{4}$O$_{16}$]$^{20-}$ groups consist of four edge-sharing [FeO$_{6}$]  distorted octrahedra. All iron ions possess a +3 oxidation state.  More detailed descriptions of the crystal structure are found elsewhere.\cite{JACS.128.15469}

The FeTe$_{2}$O$_{5}$Br compound exhibits nearly-collinear (7$^{\circ}$ of canting between the Fe1 and Fe2 sites) incommensurate antiferromagnetic ordering of its moments below T$_{N}$ = 10.6~K.\cite{PRL.103.147202,JPCS.211.012002}  The amplitude-modulated magnetic order is described with the wave vector \textbf{q}=(1/2,0.463,0).  Simultaneously, a ferroelectric polarization is induced perpendicular to \textbf{q} and the Fe$^{3+}$ moments.\cite{PRL.103.147202}  The ferroelectric order is attributed to the highly polarizable Te$^{4+}$ lone pair electrons.  Single-crystal x-ray diffraction measurements in the ordered state did not detect any change in crystal symmetry from the high temperature phase; however, clearly distinguishable changes in Fe-Te and Fe-O interatomic distances were observed that signified exchange striction as the means by which the inversion center is removed, thus allowing for ferroelectricity to arise.\cite{JPCS.211.012002}  Exchange striction is the mechanism by which magnetic ions shift away from their centrosymmetric positions to maximize their exchange interaction energies.\cite{NatMater.6.13,Nature.426.55}     

\section{Experimental Procedures}
Single crystals of FeTe$_{2}$O$_{5}$Br have been grown by standard chemical vapor phase method. Mixtures of analytical grade purity Fe$_{2}$O$_{3}$, TeO$_{2}$ and FeBr$_{3}$ powder in molar ratio 1:6:1 were sealed in the quartz tubes with electronic grade HBr as the transport gas for the crystal growth. The ampoules were then placed horizontally into a tubular two-zone furnace and heated very slowly by 50$^{\circ}$C/h to 500$^{\circ}$C. The optimum temperatures at the source and deposition zones for the growth of single crystals have been 500$^{\circ}$C and 440$^{\circ}$C, respectively. After six weeks, many dark yellow, almost orange FeTe$_{2}$O$_{5}$Br crystals with a maximum size of $8\times12\times1$~mm$^3$ were obtained. The powder x-ray diffraction pattern obtained by using a Rigaku x-ray diffractometer shows the monoclinic space group P2$_{1}$/{\it c\/} for all FeTe$_{2}$O$_{5}$Br crystals.

The zero-field temperature-dependent (5--300~K) reflectance and transmittance measurements were performed on a $1.3\times8\times6$~mm$^3$ single crystal (crystal 1) using a Bruker 113v Fourier transform interferometer in conjunction with a 4.2 K silicon bolometer detector in the spectral range 25--700 cm$^{-1}$ and a 300 K DTGS detector from 700--7,000 cm$^{-1}$.  The crystal was cooled using a flow cryostat.  Room temperature measurements from 7,000--33,000 cm$^{-1}$ were obtained with a Zeiss microscope photometer.  Appropriate polarizers were employed to span the desired spectral range.  

Field-dependent transmission measurements in the far infrared at a resolution of 0.25 cm$^{-1}$ were performed on a $0.3\times8\times5$~mm$^3$ single crystal (crystal 2) at beam line U4IR of the National Synchrotron Light Source, Brookhaven National Laboratory.  The measurements employed a Bruker IFS 66-v/S spectrometer in conjunction with a 10~T Oxford superconducting magnet and a 1.8 K silicon bolometer detector.  A free standing wire grid polarizer was oriented along the preferential polarization direction of the synchrotron beam.  

At this point it is important to establish with confidence that we cooled our single crystal below 10 K in both the Oxford superconducting magnet and the Janis flow cryostat.  The superconducting magnet is equipped with a variable temperature insert that cools in liquid He vapors.  This same system has been used to accurately measure the properties of superconducting materials down to 2~K.\cite{PRL.105.257006}  The accuracy of the cooling capacity of the flow cryostat was corroborated by comparing ratios of $\mathcal{T}_s$/$\mathcal{T}_n$ for the superconductor Nb$_{0.5}$Ti$_{0.5}$N with those same ratios measured in the superconducting magnet.  With the superconductor fastened to the cold finger of our flow cryostat with a copper clamp, the cooling capabilities, below 10 K, agreed to within $\pm$1~K. 

In materials where the crystal symmetry is lower than cubic, electrical excitations strongly depend on the orientation of the electric field.  For these low symmetry crystals, the dielectric constant is a $3\times3$ tensor and will generally have off diagonal components.  For every crystal symmetry, however, there exists a set of orthogonal Cartesian axes, called the principal dielectric axes, which diagonalize the dielectric tensor.\cite{Born&Wolf}  In an optics experiment, the principal dielectric axes are found by rotating the orientation of the sample which is placed between two polarizers that are crossed and remain fixed.  Throughout a revolution of 360$^{\circ}$, four orthogonal orientations will result in zero transmitted light beyond the second polarizer.  These four dark orientations correspond to the electric field of the incoming light aligning itself parallel to a principal dielectric axis of the system.

For a monoclinic system, the crystallographic {\it b\/} axis is unique because it is perpendicular to both the {\it a\/} and {\it c\/} axes.  Accordingly, one principal dielectric axis of the system is fixed along the crystallographic {\it b\/} axis, but the other two axes can orient themselves arbitrarily as long as they remain orthogonal.  Consequently, we performed single crystal x-ray diffraction on crystal 1 with knowledge of how the principal dielectric axes fell in relation to the crystal facets in order to determine their orientation with respect to the {\it a\/} and {\it c\/} crystallographic directions. 

X-ray diffraction measurements were collected at 300~K on a four-circle 
Panalytical X'Pert MRD in a parallel beam geometry with K$\alpha$ radiation.  A $\theta$-2$\theta$ scan was used to check that the sample surface corresponded to the {\it bc\/} plane while pole figure measurements were performed to find the spatial orientation of the {\it a\/} axis with respect to the {\it bc\/} plane.

\section{Results and Analysis}
\subsection{X-Ray diffraction}
\begin{figure}
\includegraphics[width=3.375 in]{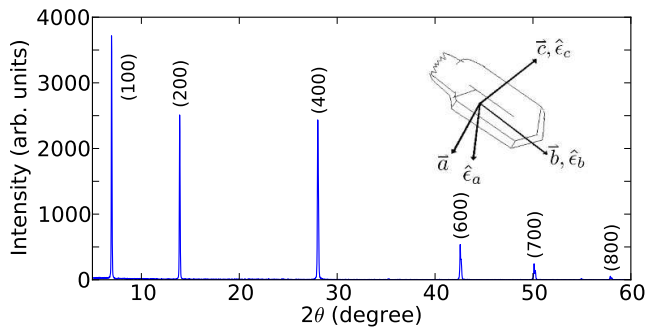}%
\caption{The XRD pattern for the sample surface aligned perpendicular to the diffraction plane. The inset depicts the orientation of the principal dielectric axes and crystallographic axes with respect to the facets of crystal 1.}
\label{XRAY}
\end{figure}

The diffraction experiment showed that the crystallographic {\it c\/} axis also coincides with a principal dielectric axis of the system.  Furthermore, the third principal dielectric axis makes an 18$^{\circ}$ angle with the crystallographic {\it a\/} axis.  The XRD patterns acquired from the sample with the surface aligned perpendicular to the diffraction plane are displayed in Fig.~\ref{XRAY}. Only very narrow (h00) diffraction lines were observed, confirming that the sample was a high quality single crystal, with the surface aligned parallel to the {\it bc\/} plane. Several acquired pole figures helped identify the orientation of the {\it a\/} axis with respect to the {\it bc\/} plane.  The inset to Fig.~\ref{XRAY} contains the results of our XRD investigation, namely, a sketch of crystal 1 with the principal dielectric axes ($\hat e_{A}$, $\hat e_{B}$, and $\hat e_{C}$) superimposed on the monoclinic crystal axes.

\subsection{Reflectance and Transmittance Spectrum}
\begin{SCfigure*}
\includegraphics[width=0.75\textwidth]{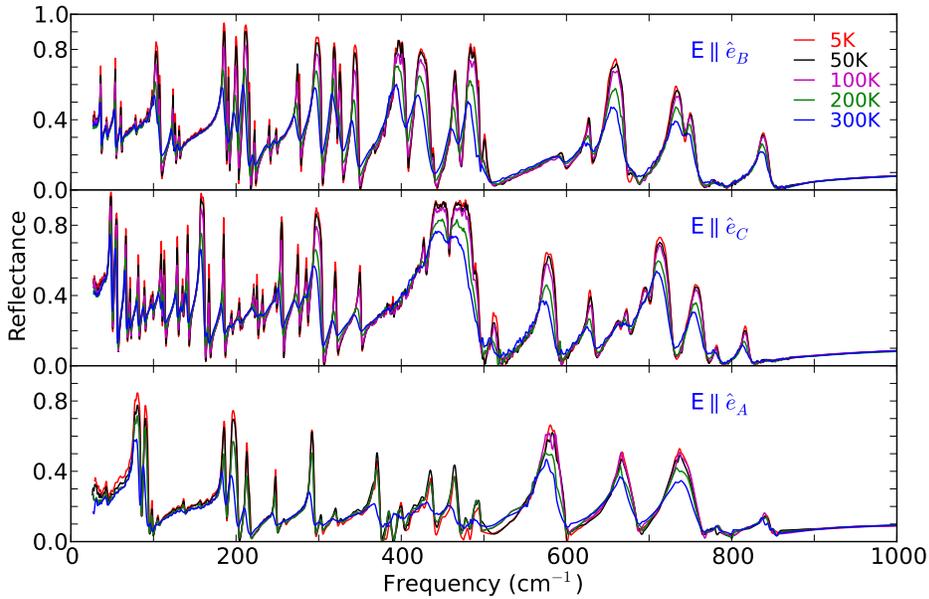}%
\caption{(Color online) Temperature-dependent reflectance spectrum of FeTe$_{2}$O$_{5}$Br along $\hat e_{A}$, $\hat e_{B}$, and $\hat e_{C}$.}   
\label{Refl_temps}
\end{SCfigure*}

The temperature-dependent reflectance spectrum of FeTe$_{2}$O$_{5}$Br for electric field oriented parallel $\hat e_{A}$, $\hat e_{B}$, and $\hat e_{C}$ in the frequency interval 30--1,000~cm$^{-1}$ (4--120~meV) is shown in Fig.~\ref{Refl_temps}.  A strong sharpening of many modes accompanied by a hardening of resonance frequencies is observed with decreasing temperature.  No drastic deviations from the room-temperature spectrum were observed along any of the three directions when cooling to 5~K.  Particular attention was given to the excitation spectrum slightly above and below the 10.6~K multiferroic ordering temperature.  The absence of anomalies in the spectra upon cooling below 10.6~K gives strong support to the observation of no crystal symmetry change in the ordered state; however, a subtle contradiction exists.\cite{PRL.103.147202}  The high temperature P2$_{1}$/{\it c\/} space group is centrosymmetric, but ferroelectricity, for which non centrosymmetry is a strict requirement, is observed below T$_N$.  This in turn suggests that new infrared modes should accompany the transition.  However, the small value of polarization measured (8.5 $\mu$C/m$^2$) leads one to suspet that a subtle transition occurs and additional infrared modes are likely below our experimental resolution ($\textless$0.5 cm$^{-1}$).   

\begin{figure}[h]
\includegraphics[width=0.5\textwidth]{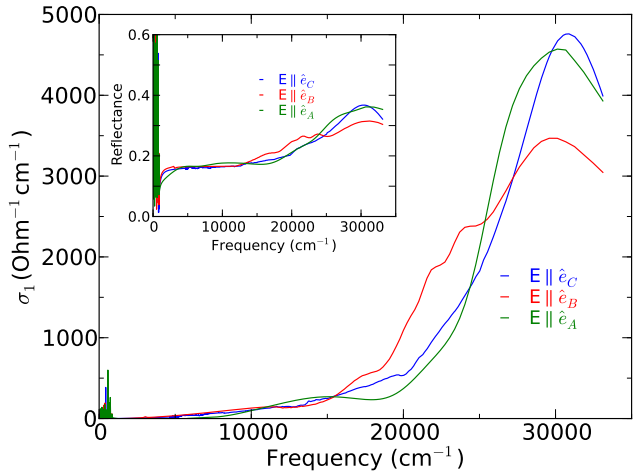}%
\caption{(Color online)  The optical conductivity in the frequency interval 30--33,000 cm$^{-1}$.  The inset shows the 300 K broadband reflectance out to 33,000 cm$^{-1}$ from which the optical conductivity was extracted via Kramers-Kronig relations.}  
\label{high_freq}
\end{figure}

Figure~\ref{high_freq} displays the optical conductivity along the three principal dielectric axes in the frequency interval 30--33,000 cm$^{-1}$.  The strong anisotropy observed in the far infrared persists at high frequencies resulting in three distinct absorption edges.  The inset of Fig.~\ref{high_freq} shows the 300 K reflectance up to 33,000 cm$^{-1}$ along $\hat e_{A}$, $\hat e_{B}$, and $\hat e_{C}$.  

\begin{figure}[htp]
\includegraphics[width=0.5\textwidth]{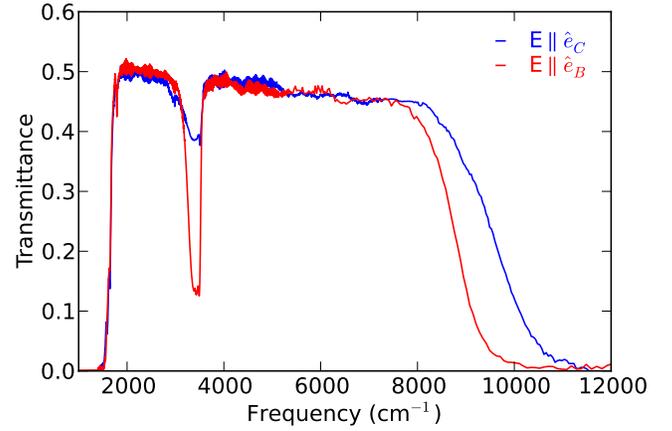}%
 \caption{(Color online) Transmittance of crystal 1 in the mid-infrared and near-infrared regions.}   
\label{Trans_temps}
\end{figure}

Crystal 1 exhibits transmission at frequencies above the infrared phonon modes and below the absorption edge.  The dimensions of the crystal only allowed for transmission with light polarized along $\hat e_{B}$ and $\hat e_{C}$.  As shown in Fig.~\ref{Trans_temps}, a sharp dip at 1850~cm$^{-1}$ as well as a broad absorption centered at 3250~cm$^{-1}$ are present along both polarizations.  Structures at these particular frequencies call to mind the absorption bands of H$_{2}$O, which in turn suggest water of hydration in the crystal structure.  But the anisotropic strengths of these absorptions can be used to rule out H$_{2}$O.  Therefore the structures suggest absorptions that are intrinsic to FeTe$_{2}$O$_{5}$Br.  The differences in energy of the high frequency transmission edge along $\hat e_{B}$ and $\hat e_{C}$ reinforce the anisotropy seen in the optical conductivity at high frequencies (Fig.~\ref{high_freq}).

\subsection{Field-dependent transmittance}
Crystal 2 transmits at frequencies below the strong infrared phonons, in the gaps in-between the phonons, and in the mid infrared between the highest phonon and the onset of interband transitions.  Particular attention was given to the transmission below the strong infrared phonons.  No appreciable change in the spectra were observed for transmittance at 5~K measured along $\hat e_{B}$ and $\hat e_{C}$ with magnetic fields of 10~T applied along all three crystallographic directions.  A class of low energy magnetic excitations called antiferromagnetic resonance modes (AFMR) are predicted to exist at finite frequencies in zero field and shift upon the application of external fields.  AFMR modes in the FeTe$_{2}$O$_{5}$Br have been reported in a recent electron spin resonance study below 400~GHz ($\sim$13~cm$^{-1}$).\cite{PRB.86.054402}  However, in an external field of 10~T the modes remain slightly below our measurable frequency range (15~cm$^{-1}$).

\subsection{Determination of optical properties}

We have used the measured reflectance and transmittance spectra of crystal 1 to extract the complex dielectric function.  Because Kramers-Kronig analysis requires a single-bounce reflectance spectrum over a broad frequency region, we must first correct the measured reflectance spectrum in regions where the crystal transmits by using a combined reflection and transmission analysis.  An attempt to solve for the single bounce reflectance by inverting the full reflection and transmission equations leads to a transcendental equation with infinitely many solutions.  Nevertheless, using the method employed by Zibold {\it et al.\/},\cite{PRB.53.11734} i.e., assuming that {\it k\/} is small in the region of interest, leads to an approximation of reflection and transmission that can be solved numerically for a single solution.  

\begin{figure}[htp]
\includegraphics[width=0.5\textwidth]{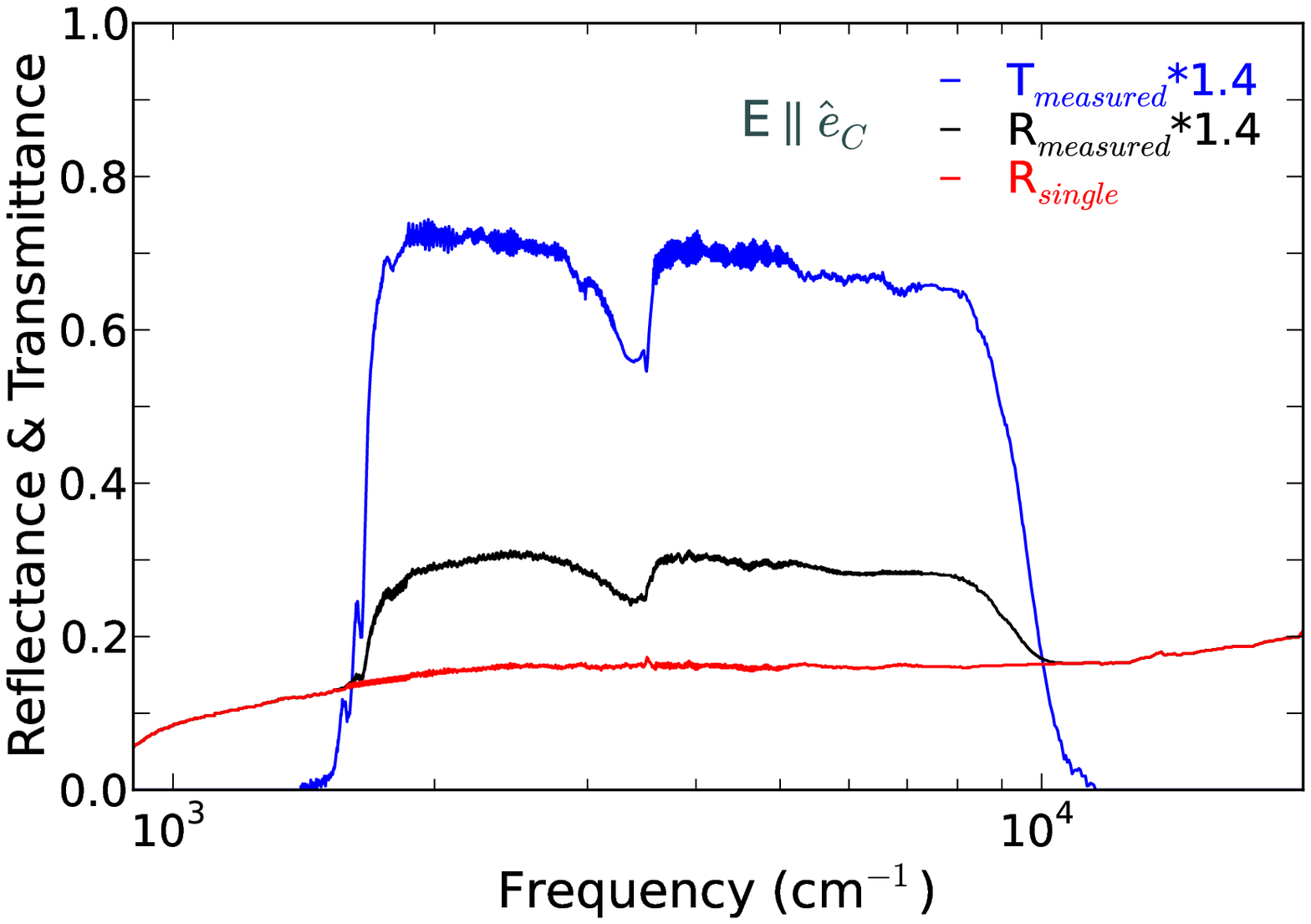}%
 \caption{(Color online) Single-bounce reflectance (red) as approximated from the scaled measured reflectance (black) and the scaled measured transmittance (blue) of crystal 1.  The elevated level of measured reflectance throughout the transmission interval is expected due to additional light reflected from the back surface of the crystal.}   
\label{R_s}
\end{figure}

The measured reflectance (not shown) and transmittance (Fig.~\ref{Trans_temps}) in the frequency interval 1500--10000~cm$^{-1}$ were significantly lower than expected throughout this region of low absorption (Fig.~\ref{high_freq}).  The low measured values were attributed to the scattering of light from facets existing on the back surface of crystal 1.  To compensate, we assumed a mask on the back surface of crystal 1 which lead to scaling the measured reflectance and transmittance up by a factor of 1.4 in the interval 1500--10000~cm$^{-1}$, thus explaining the discrepancy between Fig.~\ref{Trans_temps} and Fig.~\ref{R_s}.  The scaled reflectance and transmittance yielded a single bounce reflectance which agreed well with the measured bulk reflectance in regions where crystal 1 did not transmit, as shown in Fig.~\ref{R_s}.  
 
Kramers-Kronig analysis can be used to estimate the real and imaginary parts of the dielectric function from the bulk reflectance $R(\omega$).\cite{Wooten}  
Before calculating the Kramers-Kronig integral, the low frequency data (0.1--30~cm$^{-1}$) were extrapolated using parameters from the complex dielectric function model described in Sec.~III~E.  At high frequencies ($8 \times 10^{4}$--$2.5 \times 10^{8}$~cm$^{-1}$) the reflectance was approximated from known x-ray scattering functions of the constituent atoms.  The gap between our highest measurable frequency and the x-ray reflectance data was bridged with a Lorentz oscillator at 60,000~cm$^{-1}$.  Subsequently, the phase shift on reflection was obtained via Kramers-Kronig analysis, and the optical properties were calculated from the reflectance and phase.

\subsection{Oscillator-model fits}
\begin{figure}[h]
\includegraphics[width=0.5\textwidth]{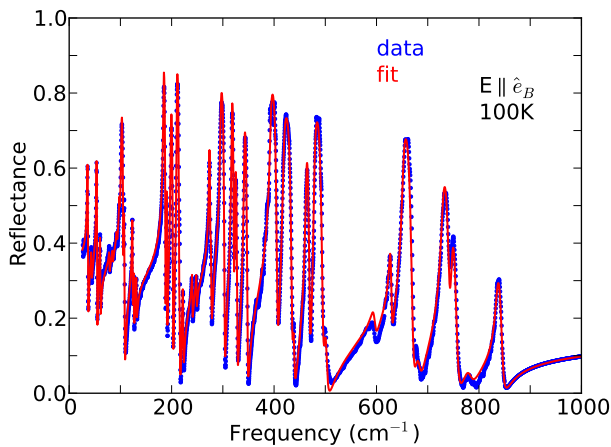}%
 \caption{(Color online) Experimental reflectance (blue points) and calculated reflectance (red line) from the Drude-Lorentz model of the FeTe$_{2}$O$_{5}$Br 100 K $\hat e_{B}$ dielectric function.}  
\label{Fit_refl_100K}
\end{figure}

The measured reflectance was fit with a Drude-Lorentz model to obtain a second estimate of the complex dielectric function in the infrared range.  The model assigns a lorentzian oscillator to each distinguishable phonon in the spectrum plus a high frequency permittivity, $\epsilon_\infty$, to address the contribution of electronic excitations.  The model has the following mathematical form:
\begin{equation}
\epsilon = \displaystyle\sum_{j=1}^{\infty}\frac{S_j\omega_j^2}{{{\omega_j}^2}-{{\omega}^2}-{i\omega\gamma_j}} +\epsilon_\infty\;\;\;,
\label{epsilonDL}
\end{equation}
where $S_j$, $\omega_j$, and $\gamma_j$ signify the oscillator strength, center frequency, and the full width at half max (FWHM) of the jth lorentzian oscillator.  The Drude-Lorentz complex dielectric function is used to calculate the reflectivity, which is compared to the original measured quantity in Fig.~\ref{Fit_refl_100K}.  Similar qualities of fits were obtained for all other temperatures and polarizations. 

\begin{table}[tr!]
\caption{Oscillator parameters for experimentally observed phonons along $\hat e_{A}$ (at 20 K).  The motions of atoms associated with groups of phonons have been assigned when the pattern was apparent.  The atoms are listed in descending order of their calculated net atomic displacement.}
\begin{ruledtabular}
\begin{tabular}{l*{6}{c}r}
Idx & Osc Str & Ctr Freq & FWHM & Degen & Asgn \\
 & S & $\omega$ (cm$^{-1}$) & $\gamma$ (cm$^{-1}$) & idx in $\hat e_{C}$ &  \\
\hline
  1 & 0.207 & 45.5 & 2.0 & 1 & Br \\
  2 & 1.320 & 79.6 & 2.1 & 5 & - \\
  3 & 0.356 & 88.4 & 1.1 & 6 & Br,Te,O,Fe \\
  4 & 0.042 & 98.1 & 1.0 & 8 & Br,Te,O,Fe \\
  5 & 0.056 & 107.5 & 2.1 & 9 & Br,Te,O,Fe \\
  6 & 0.048 & 137.1 & 1.9 & 12 & Br,Te,O,Fe \\
  7 & 0.015 & 157.6 & 1.6 & 14 & Te,Br,O,Fe \\
  8 & 0.023 & 166.0 & 2.2 & 15 & Te,Br,O,Fe \\
  9 & 0.354 & 184.9 & 1.3 & 16 & Te,O,Fe \\
  10 & 0.519 & 194.3 & 2.4 & 17 & Te,O,Fe \\
  11 & 0.159 & 211.7 & 2.0 &  & Te,O,Fe \\
  12 & 0.007 & 236.4 & 4.9 & 22 & - \\
  13 & 0.084 & 247.8 & 1.7 & 23 & - \\
  14 & 0.009 & 256.5 & 4.0 & 24 & - \\
  15 & 0.169 & 291.9 & 1.8 & 27 & Fe,O,Te \\
  16 & 0.030 & 302.8 & 3.4 &  & Fe,O,Te \\
  17 & 0.033 & 321.6 & 3.6 & 28 & Fe,O,Te \\
  18 & 0.006 & 328.0 & 2.4 &  & Fe,O,Te \\
  19 & 0.003 & 348.5 & 2.0 & 29 & Fe,O,Te \\
  20 & 0.102 & 370.0 & 2.5 &  & - \\
  21 & 0.008 & 387.1 & 2.6 &  & - \\
  22 & 0.030 & 401.3 & 4.1 &  & - \\
  23 & 0.005 & 418.6 & 4.0 &  & - \\
  24 & 0.005 & 426.2 & 2.6 & 30 & - \\
  25 & 0.090 & 435.0 & 3.9 & 31 & - \\
  26 & 0.119 & 463.6 & 5.4 & 32 & O,Fe,Te \\
  27 & 0.004 & 481.0 & 1.4 &  & O,Fe,Te \\
  28 & 0.029 & 493.2 & 4.9 & 33 & O,Fe,Te \\
  29 & 0.350 & 575.3 & 7.5 & 35 & O,Fe \\
  30 & 0.203 & 666.1 & 8.0 & 37 & O,Fe \\
  31 & 0.257 & 730.5 & 13.9 &  & O,Fe \\
  32 & 0.018 & 784.0 & 9.2 & 42 & - \\
\end{tabular}
\end{ruledtabular}
\label{epsA}
\end{table} 
\begin{table}[tr!]
\caption{Oscillator parameters for experimentally observed phonons along $\hat e_{C}$ (at 20 K).  The motions of atoms associated with groups of phonons have been assigned when the pattern was apparent.  The atoms are listed in descending order of their calculated net atomic displacement.}
\begin{ruledtabular}
\begin{tabular}{l*{6}{c}r}
Idx & Osc Str & Ctr Freq & FWHM & Degen & Asgn \\
 & S & $\omega$ (cm$^{-1}$) & $\gamma$ (cm$^{-1}$) & idx in $\hat e_{A}$ &  \\
\hline
  1 & 4.275 & 46.7 & 0.2 & 1 & Br \\
  2 & 1.335 & 53.9 & 0.5 &  & Br \\
  3 & 0.951 & 65.8 & 1.0 &  & - \\
  4 & 0.216 & 71.7 & 0.8 &  & - \\
  5 & 0.435 & 81.2 & 0.9 & 2 & - \\
  6 & 0.205 & 89.3 & 0.9 & 3 & Br,Te,O,Fe \\
  7 & 0.061 & 97.0 & 0.8 &  & Br,Te,O,Fe \\
  8 & 0.130 & 99.0 & 1.1 & 4 & Br,Te,O,Fe \\
  9 & 0.533 & 109.2 & 0.8 & 5 & Br,Te,O,Fe \\
  10 & 0.294 & 112.8 & 1.0 &  & Br,Te,O,Fe \\
  11 & 0.369 & 128.5 & 0.8 &  & Br,Te,O,Fe \\
  12 & 0.127 & 135.4 & 0.9 & 6 & Br,Te,O,Fe \\
  13 & 0.411 & 140.6 & 0.9 &  & - \\
  14 & 0.857 & 157.0 & 0.5 & 7 & Te,Br,O,Fe \\
  15 & 0.107 & 166.8 & 1.0 & 8 & Te,Br,O,Fe \\
  16 & 0.212 & 184.5 & 0.6 & 9 & Te,O,Fe \\
  17 & 0.039 & 194.8 & 1.2 & 10 & Te,O,Fe \\
  18 & 0.066 & 204.0 & 1.4 &  & - \\
  19 & 0.014 & 206.3 & 1.1 &  & - \\
  20 & 0.085 & 221.7 & 1.1 &  & - \\
  21 & 0.101 & 225.4 & 0.9 &  & - \\
  22 & 0.065 & 232.6 & 1.0 & 12 & - \\
  23 & 0.107 & 248.0 & 1.5 & 13 & - \\
  24 & 0.291 & 254.2 & 0.6 & 14 & - \\
  25 & 0.239 & 274.4 & 1.1 &  & - \\
  26 & 0.195 & 284.9 & 1.2 &  & - \\
  27 & 0.633 & 294.8 & 1.4 & 15 & Fe,O,Te \\
  28 & 0.138 & 319.0 & 1.9 & 17 & Fe,O,Te \\
  29 & 0.173 & 349.0 & 2.2 & 19 & Fe,O,Te \\
  30 & 0.514 & 426.7 & 2.4 & 24 & - \\
  31 & 1.562 & 437.0 & 2.3 & 25 & - \\
  32 & 0.088 & 459.3 & 3.4 & 26 & O,Fe,Te \\
  33 & 0.006 & 488.5 & 4.3 & 28 & O,Fe,Te \\
  34 & 0.067 & 509.6 & 9.6 &  & O,Fe,Te \\
  35 & 0.260 & 573.5 & 5.6 & 29 & O,Fe \\
  36 & 0.096 & 626.7 & 5.2 &  & O,Fe \\
  37 & 0.015 & 662.0 & 3.2 & 30 & O,Fe \\
  38 & 0.061 & 672.0 & 5.7 &  & O,Fe \\
  39 & 0.182 & 693.5 & 8.9 &  & O,Fe \\
  40 & 0.172 & 707.4 & 5.2 &  & O,Fe \\
  41 & 0.101 & 749.8 & 9.4 &  & - \\
  42 & 0.014 & 780.4 & 6.6 & 32 & - \\
  43 & 0.033 & 814.7 & 6.9 &  & - \\
\end{tabular}
\end{ruledtabular}
\label{epsC}
\end{table} 
\begin{table}[tr!]
\caption{Oscillator parameters for experimentally observed phonons along $\hat e_{B}$ (at 20 K).  The motions of atoms associated with groups of phonons have been assigned when the pattern was apparent.  The atoms are listed in descending order of their calculated net atomic displacement.}
\begin{ruledtabular}
\begin{tabular}{l*{5}{c}r}
Idx & Osc Str & Ctr Freq & FWHM & Asgn \\
 & S & $\omega$ (cm$^{-1}$) & $\gamma$ (cm$^{-1}$) &  \\
\hline
  1 & 1.388 & 35.7 & 0.8 & Br \\
  2 & 0.330 & 43.5 & 1.4 & Br \\
  3 & 0.899 & 53.2 & 0.8 & - \\
  4 & 0.289 & 60.5 & 0.9 & Br,Te,O,Fe \\
  5 & 0.119 & 79.1 & 1.3 & Br,Te,O,Fe \\
  6 & 0.075 & 85.0 & 1.2 & Br,Te,O,Fe \\
  7 & 0.052 & 91.2 & 1.0 & Br,Te,O,Fe \\
  8 & 0.138 & 93.3 & 1.3 & - \\
  9 & 0.259 & 98.0 & 0.9 & - \\
  10 & 1.359 & 102.0 & 1.1 & - \\
  11 & 0.073 & 106.9 & 0.7 & - \\
  12 & 0.170 & 123.7 & 0.7 & Te,Br,O,Fe \\
  13 & 0.092 & 126.3 & 0.9 & Te,Br,O,Fe \\
  14 & 0.074 & 131.4 & 1.1 & Te,Br,O,Fe \\
  15 & 1.066 & 184.0 & 0.4 & Te,O,Fe \\
  16 & 0.128 & 190.2 & 0.7 & Te,O,Fe \\
  17 & 0.491 & 198.6 & 0.8 & Te,O,Fe \\
  18 & 0.440 & 210.0 & 0.3 & - \\
  19 & 0.023 & 215.3 & 1.3 & - \\
  20 & 0.053 & 222.0 & 1.1 & - \\
  21 & 0.058 & 240.1 & 1.3 & - \\
  22 & 0.028 & 249.0 & 0.9 & - \\
  23 & 0.274 & 273.9 & 0.9 & Fe,O,Te \\
  24 & 0.102 & 275.8 & 1.4 & Fe,O,Te \\
  25 & 0.661 & 295.5 & 1.3 & Fe,O,Te \\
  26 & 0.366 & 317.6 & 1.1 & Fe,O,Te \\
  27 & 0.101 & 324.3 & 1.5 & Fe,O,Te \\
  28 & 0.261 & 341.7 & 1.2 & Fe,O,Te \\
  29 & 0.808 & 393.7 & 2.6 & Fe,O,Te \\
  30 & 0.371 & 418.0 & 4.1 & - \\
  31 & 0.009 & 437.7 & 3.5 & - \\
  32 & 0.209 & 462.8 & 3.2 & O,Fe,Te \\
  33 & 0.173 & 478.0 & 3.3 & O,Fe,Te \\
  34 & 0.012 & 499.0 & 4.6 & - \\
  35 & 0.009 & 508.3 & 1.2 & - \\
  36 & 0.047 & 595.3 & 6.9 & O,Fe \\
  37 & 0.080 & 626.6 & 3.6 & O,Fe \\
  38 & 0.323 & 652.0 & 4.9 & O,Fe \\
  39 & 0.018 & 682.1 & 8.2 & O,Fe \\
  40 & 0.217 & 728.1 & 7.8 & O,Fe \\
  41 & 0.027 & 746.2 & 7.4 & - \\
  42 & 0.031 & 779.9 & 23.5 & - \\
  43 & 0.071 & 836.0 & 9.6 & - \\
\end{tabular}
\end{ruledtabular}
\label{epsB}
\end{table} 

 \section{Discussion}
\subsection{Group theory and lattice dynamics}
Due to the lack of anomalies upon cooling below 10.6~K, we turn our focus to the highly anisotropic far-infrared excitation spectrum.  The number of phonon modes expected in FeTe$_{2}$O$_{5}$Br can be found by group-theory analysis.  Using the SMODES program,\cite{SMODES} we arrive at the following distribution of modes:

\begin{equation}
\Gamma^{optical} = 54A_{\it~g}^{\it(R)}+54B_{\it~g}^{\it(R)}+53A_{\it~u}^{\it(IR)}+52B_{\it~u}^{\it(IR)}\;\;\;,
\label{Gamma_optical}
\end{equation}
where (R) and (IR) denote Raman active and infrared active modes respectively.  The 53$A_{\it u}$ modes are expected to lie along the unique {\it b\/} axis ($\hat e_{B}$) of the monoclinic system.  The 52$B_{\it u}$ modes are expected to lie in the {\it ac\/} plane, and one should be able to resolve all 52 modes using the two orthogonal polarization spectra measured in this plane.  However, the merging of weaker modes near stronger modes can hinder this count.  We therefore employed lattice dynamical calculations to determine the relative resonance frequency, mode intensity, as well as displacement pattern for the 105 infrared active modes.  The calculations, which utilized the structure and valence as reported by Becker {\it et al.\/},\cite{JACS.128.15469} were based on a real-space summation of screened coulomb interactions involving a spherical cut-off boundary.\cite{JPC.110.8254} The following two sections describe the method used to compare the observed and calculated spectrum for the $A_{\it u}$ and $B_{\it u}$ modes respectively.

\subsection{52$B_{\it u}$ modes}
All 52$B_{\it u}$ modes predicted lie in the {\it ac\/} plane and should possess a finite intensity along both measured polarizations in this plane unless the direction of the induced dipole moment for a particular vibration is parallel to either $\hat e_{A}$ or $\hat e_{C}$.  (Small angle deviations from the parallel case may be hard to detect.)  For all other modes one must sort out which excitations along $\hat e_{A}$ and $\hat e_{C}$ corresponding to the same physical vibration so that they are not counted twice.  To count the modes, we used the Drude-Lorentz parameters for each phonon at 20~K as well as the resonant frequency and mode intensity from our lattice dynamical calculation.  With knowledge of the optical conductivity spectra along $\hat e_{A}$ and $\hat e_{C}$ at 20~K, modes were grouped together if they were observed in the same small frequency interval (on the order of a typical linewidth) of one another in the two spectra.  Next, we compared the experimental linewidths of the grouped modes to ensure that they would overlap if plotted side-by-side. As two final criteria, we compared both the ratio of experimental oscillator strengths with the computed oscillator strengths, and also compared the displacement patterns generated by our calculation to make sure they were in good agreement as to which excitations corresponded to the same physical vibrations.  All 52$B_{\it u}$ modes predicted in the {\it ac\/} plane were observed in the experimental spectrum.  Lorentz oscillator parameters for all modes as well as modes present along both $\hat e_{A}$ and $\hat e_{C}$ polarizations can be found in Table~\ref{epsA} and Table ~\ref{epsC}.

\subsection{53$A_{\it u}$ modes}
\begin{figure}[h]
\includegraphics[width=0.5\textwidth]{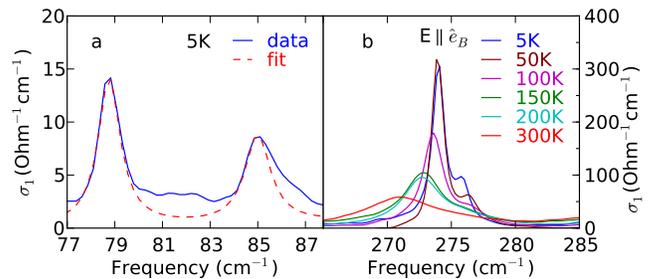}%
\caption{(Color online)  The left panel (a) shows unaccounted for spectral weight on the high frequency side of the asymmetric oscillator at 85 cm$^{-1}$ along with a fit to a symmetric oscillator (79 cm$^{-1}$).  We suspect the asymmetric shoulder is the result of a buried mode.  The right panel (b) depicts the appearance of a once buried mode ($\sim$276~cm$^{-1}$) upon cooling.}  
\label{Buried}
\end{figure}

Counting the experimentally observed $A_{\it u}$ modes is straightforward; we observe 43 of the 53 predicted modes along this direction.  We believe that 10 modes were not resolved as a result of weaker modes being buried within stronger modes and merging together in the spectrum.  In the optical conductivity, all single phonon excitations should be represented by symmetric lorentzian oscillators.  Figure~\ref{Buried} left panel depicts a slightly asymmetric lorentzian oscillator with a resonance of 85 cm$^{-1}$.  Our attempt to fit the resonance at 85 cm$^{-1}$ with a symmetric lorentzian oscillator yields unaccounted for spectral weight on the high frequency side of the resonance.  A symmetric lorentzian oscillator centered at 79 cm$^{-1}$ and corresponding fit are also included as a reference.  (Note that the additional spectral weight between the resonances at 75 cm$^{-1}$ and 85 cm$^{-1}$ is attributed to absorption between the bands and therefore not to be mistaken for a lattice vibration).  We suspect that this asymmetric shoulder is due to a weak buried mode that is not fully resolved.  To further support our interpretation of buried modes, the right panel of Fig.~\ref{Buried} depicts the appearance of a once buried mode ($\sim$276~cm$^{-1}$) upon cooling.  The Lorentz oscillator parameters of the 52$B_{\it u}$ modes at 20~K are found in Table~\ref{epsB}.

\subsection{Assignment of Modes}
The assignment of a particular phonon, observed in the experimental spectrum, to its precise atomic displacement pattern cannot be made without sacrificing objective certainty because of the density of the phonon spectrum.    Assignments to groups of modes, however, can be made through the displacement patterns generated by our lattice dynamical calculations. Using a simplified model of a phonon as a driven oscillator with resonant frequency inversely proportional to mass, we proceed to list the constituent atoms in the structure from heaviest to lightest: Te, Br, Fe, and O.  Normally, the low frequency modes involve the heaviest atoms, but in the FeTe$_{2}$O$_{5}$Br structure the Br atoms have a weak interaction with the other atoms and therefore are responsible for the lowest frequency phonons.  At a certain frequency the motion of Br atoms will cease.  Likewise, at a higher frequency the vibration of atoms involving Te will cease, leaving the motion of Fe and O to be expected in the highest frequency phonons.   Accordingly, the phonon spectrum can be divided into three clusters that is reflected in Fig.~\ref{Refl_temps} (namely, below 180 cm$^{-1}$, between 180 cm$^{-1}$ and 530 cm$^{-1}$, and above 530 cm$^{-1}$).  The clustering is also observed in the calculated spectrum via the displacement patterns produced, in terms of first the Br and subsequently the Te ceasing to vibrate.    Within each cluster trends are to be noted on the following principle: atoms are listed corresponding to their net atomic displacement. This approach, in turn, has allowed us to identify another characteristic of atomic displacement within the three clusters.  The atom of focus in a cluster (e.g., Br in the first cluster) has a net atomic displacement ever diminishing. Cessation does not arrive abruptly. The three clusters are profiled in each of Table~\ref{epsB}, Table~\ref{epsA}, and Table~\ref{epsC}.

Apropos the third cluster that focuses on Fe and O, an additional resource is provided by the familiar orthoferrite family of compounds because the FeO$_{6}$ octrahedral building block that is present in FeTe$_{2}$O$_{5}$Br is itself a defining unit of the orthoferrite family.  Given the abundant literature on vibrational assignments made to the FeO$_{6}$ ochrahedral unit,\cite{PRB.75.220102,JMMM.270.380} we are consequently able to map these well recognized assignments onto our experimental findings, thus corroborating both perspectives.  

\section{Conclusions}
Far-infrared measurements on multiferroic single-crystal FeTe$_{2}$O$_{5}$Br reveal no drastic anomalies in the excitation spectrum below the multiferroic ordering temperature.  Nonetheless, a thorough inspection of the phonon spectra has lead to the identification of all 52 predicted modes in the degenerate {\it ac\/} plane and 43 of the 53 predicted modes along the unique {\it b\/} axis.  The absence of 10 modes along the {\it b\/} axis is presumably due to the merging of weaker modes near stronger modes in the dense excitation spectrum. 

The motions of individual atoms have been assigned to groups of phonons via the displacement patterns produced through our lattice dynamical calculation.    The assignments divide the overall phonon spectrum into three clusters, which are characterized by an atom that is common to every vibration in that cluster.   We observed that as frequency increases, vibrations involving the motion of Br atoms will progressively cease.  Likewise, at higher frequencies the vibrations involving Te atoms will cease in the same manner, leaving the motion of Fe and O to be expected in the highest frequency phonons.  Moreover, the particular atoms involved in a vibration are listed in descending order of net atomic displacement, which in turn makes clear a trend within each cluster (namely, the atom of focus in a particular cluster has a net atomic displacement ever diminishing).  
 
It remains to compare our results to a recent infrared study by Pfuner {\it et al.\/}\cite{JCMP.21.375401} on the similar FeTe$_{2}$O$_{5}$Cl compound where an equal number of phonons were predicted.  Our analysis of the phonon spectrum agrees with the infrared study of Pfuner {\it et al.\/} in that we also believe modes resonating at equivalent energy scales can overlap, preventing them from being resolved separately, but we differ on other issues.  Pfuner {\it et al.\/} report a combined 38 modes along the $\hat e_{B}$ and $\hat e_{C}$ directions. We observe more modes along the unique {\it b\/} axis alone, as well as modes below 50 cm$^{-1}$, whereas the study by Pfuner {\it et al.\/} does not report measurements below this frequency.  

We also observe drastically different and more strongly anisotropic high frequency optical properties.  Pfuner {\it et al.\/} report an absorption edge around 4,000 cm$^{-1}$ in FeTe$_{2}$O$_{5}$Cl, where as we observe low absorption accompanied by substantial transmission through the FeTe$_{2}$O$_{5}$Br crystal in this region.  In speculating about the differences in high frequency optical properties of the two similar compounds, we point out that spurious changes of slope of the reflectance throughout the mid-infrared region, such as those resulting from the contribution of back surface reflection (black line in Fig.~\ref{R_s}), can drastically change the Kramers-Kronig-derived optical properties.  

Transmission measurements as well as reflectance along $\hat e_{A}$ were not reported in the study by Pfuner {\it et al.\/}.  Similar to the FeTe$_{2}$O$_{5}$Br system reported here, the FeTe$_{2}$O$_{5}$Cl system also showed no response to magnetic fields above or below the magnetic ordering temperature.

\begin{acknowledgments}
The authors thank M. Pregelj for providing low-temperature crystallographic information files.  This work was supported by DOE through grant DE-FG02-02ER45984 at UF and DE-AC02-98CH10886 at the NSLS.                    

\end{acknowledgments}

\end{document}